\documentclass[aip,jcp,reprint]{revtex4-1} 

\usepackage[pdftex]{graphicx}
\usepackage{amsmath}
\usepackage{color}
\usepackage{hyperref}

\begin{document}

\title{Liquids that form due to
dynamics of the molecules that depend on the local density
}

\author{Richard P. Sear}
\email{r.sear@surrey.ac.uk}

\affiliation{Department of Physics, University of Surrey, Guildford, Surrey
GU2 7XH, United Kingdom}

\begin{abstract}
RNA molecules in living cells form what look like liquid droplets formed by liquid/liquid phase separation. But unlike the molecules in conventional phase separating mixtures, RNA molecules are transported by molecular motors that consume energy and so are out of equilibrium. Motivated by this
we consider what sort of simple rules for the dynamics of model
mRNA molecules lead to liquid/liquid phase separation. We find that
dynamics that slow as the local density of molecules increases, drive
the formation of liquids. We also look at the analogous separation of the two blocks of a block copolymer, in which the monomers of one block have dynamics that depend on the local density of monomers of that block. We find that this block condenses and separates from the monomers of the other block. This is a simple model of the out-of-equilibrium domain formation found in the chromatin in the nucleus of cells.
\end{abstract}

\maketitle 


The contents of living cells are in the liquid state.
But these contents are not in one liquid phase, they appear to be
in a number of coexisting liquid phases. There are what look like
liquid droplets in both the cytoplasm and the nucleus
\cite{brangwynne09,brangwynne11sm,thomas11,sear15_cecam,buchan09,lee13,
kedersha13,misteli07,cope10,mao11,brangwynne11nuc,ganai14}. For example, in the cytoplasm mRNA molecules can undergo what
looks like liquid/liquid phase separation, to produce
droplets enriched in the mRNA molecules
\cite{brangwynne09,brangwynne11sm,thomas11,sear15_cecam,buchan09,kedersha13,lee13}.
But these droplets cannot be at thermodynamic equilibrium.
They are affected when the molecular motors dynein or kinesin
are knocked down \cite{loschi09}. 
These motors consume energy and actively move mRNA
molecules \cite{fusco03}.
This suggests
that mRNA molecules do not simply diffuse into and out
of these liquid droplets, they are actively transported into
or out of these droplets.

\begin{figure}[bth!]
\includegraphics[width=3.0cm,angle=0]{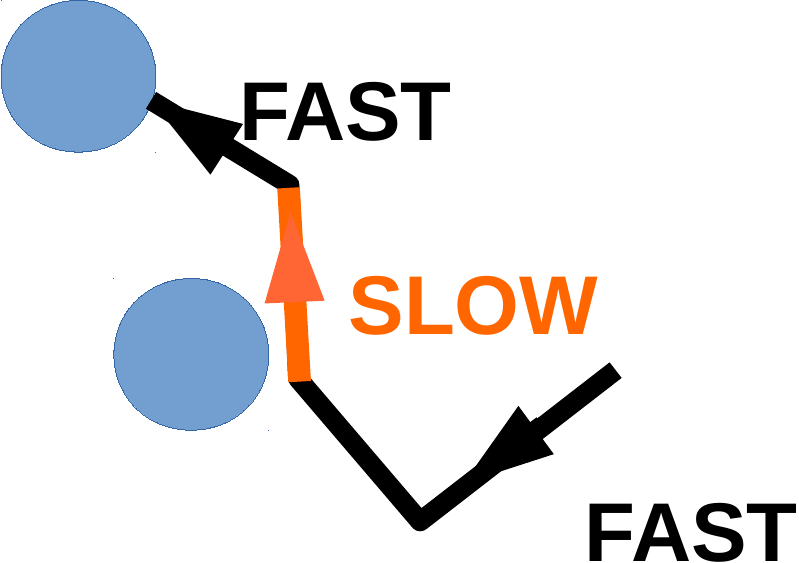}
\caption{Schematic illustrating out-of-equilibrium dynamics that depend
on the local density of molecules.
Molecules are shown as blue discs, the lines with arrows indicate the path
the top molecule has taken. It slowed down when it was near the bottom
molecule.
}
\label{fig:schem_dyn}
\end{figure}

Inspired by this, we wanted a simple model of density-dependent
dynamics that generates condensation. These dynamics should be
via hops
from one point to another; the hops do not conserve momentum, and
there is no well defined velocity.
This is our simple model of a motor translating mRNA molecules
along microtubules in a cell. 
It turns out that very simple models show condensation.
Rather generically, liquids appear whenever 
the hopping rate of a molecule decreases
as the local density of molecules increases.
These out-of-equilibrium dynamics are illustrated in a schematic in
Fig.~\ref{fig:schem_dyn}. The dynamics stabilise liquid
droplets as evaporation of molecules from a droplet's surface
is reduced by the slow hopping rate of molecules out of the dense
liquid. Attractions slow hopping rates but any non-equilibrium
mechanism that also slows hopping out of dense regions will
tend to have a similar effect, whether or not that are any
attraction energies directly involved.
The Edinburgh group and others
\cite{stenhammer13,wittkowski14,stenhammer14,stenhammer15,
redner13,mognetti13,cates14,solon14}, have extensively studied
a system with rather different microscopic dynamics; in their systems
the particles
have a well-defined velocity. However, despite the differences
in the microscopic dynamics, here we
are studying models in which the mobility decreases
as density increases, just as they have, and so we see qualitatively
very similar condensation into
liquid-like droplets.

We start with
the familiar 2D lattice gas \cite{binder_book10,chandler_book,onsager44}.
Our lattice is $L$ by $L$ lattice sites and has $N$ molecules in total.
This is a very simple model of mRNA molecules in a cell, where
we effectively integrate over all other molecules,
to allow us to explicitly consider
only these mRNA molecules. The interactions are then of mRNA
molecules in the presence of all these other molecules, and the
dynamics are those in the presence of these other molecules,
including motor proteins. As we integrate over these other
molecules liquid/liquid separation becomes condensation of the
one species we explicitly consider.

We start with
the familiar Kawasaki
dynamics \cite{binder_book10,chandler_book}.
They of course obey detailed balance.
For Kawasaki dynamics, time is measured in cycles, and in each
cycle $L^2$ lattices sites are selected at random, one after
another.
When a site is selected then one of its four neighbouring sites
is chosen at random. If one of the pair of sites is occupied
by a molecule and the other is empty, then we attempt
to move the molecule
from one site to the other.
This is done as follows. The change in the number of neighbours
of the molecule, $\Delta n$, is computed.
If $\Delta n\ge 0$ the move is always accepted
as then the energy decreases or stays the same, but if $\Delta n<0$ the move
is rejected with probability $1-\exp[\epsilon \Delta n/kT]$.
Here $\epsilon$ is an interaction energy.

When the ratio $\epsilon/kT>1.76$ \cite{onsager44,chandler_book},
a liquid phase
forms. The attractions between
the molecules cause them to condense into a liquid phase.
Droplets of liquid nucleate and grow until there is a single
large droplet coexisting with a vapour.

We now introduce a simple model of dynamics where
the rate a molecule hops at depends on the
local environment at the start position of its move.
At first sight these dynamics may look as though they violate detailed
balance, but we will show that they do not.
The dynamics are as follows.
Select a lattice site at random.
If it is occupied by a molecule, attempt to move it to a randomly
selected site within a radius $r_{M}$. If the site chosen is already
occupied the attempt is always rejected.
If the chosen site is empty the move
is made with probability $p=\exp[-\alpha_D n_n]$,
for $n_n$ the number
of neighbours of the molecule in its starting position.
Here $\alpha_D>0$ is a parameter that couples the dynamics
to the local density of molecules.
In this model the higher the local density of molecules
the slower the hopping rate of a molecule. This could be for
a number of reasons, such as neighbouring molecules inhibiting motor
transport via mechanisms such as direct binding or signalling.
Alternatively, even if motors are not involved, then
a local density-dependent slow down could be caused by
an out-of-equilibrium process such as a phosphorylation/dephosphorylation
cycle that modulates binding and so diffusion rates.

\begin{figure}[bt!]
\includegraphics[width=3.2cm,angle=0]{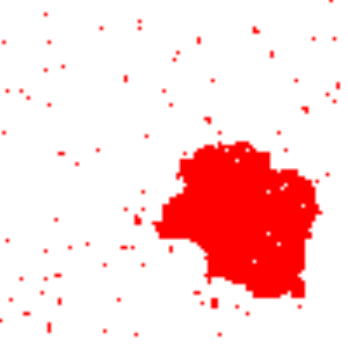}
\caption{Snapshot of configurations of systems
of $L=100$ by $L=100$ sites, with $15\%$ of the sites occupied by molecules.
This is for our out-of-equilibrium model; $\alpha_D=2.5$ and $r_M=8$,
which corresponds to the Ising lattice gas at $\epsilon/kT=2.5$ .
We have imposed periodic boundary conditions.
}
\label{fig:snaps}
\end{figure}

Our density-dependent dynamics map onto the standard Ising lattice gas,
as can be seen as follows. To do this we need
to consider the transition probability between state $i$ and state
$j$, $p_{ij}$; these states $i$ and $j$
 correspond to two positions of a molecule
on the lattice.
For transition between a state $i$ with
a molecule surrounded by $n_n(i)$ neighbours and
a state $j$ where the molecule has $n_n(j)$ neighbours, the
ratio of the
transition probabilities $i$ to $j$, and $j$ to $i$ are
\begin{equation}
\frac{p_{ij}}{p_{ji}}=\frac{\exp[-\alpha_D n_n(i)]}{\exp[-\alpha_D n_n(j)}
=\exp\left[-\alpha_D[ n_n(i)- n_n(j)]\right]
\end{equation}
These satisfy detailed balance, and map to the Ising lattice
gas with $\epsilon/kT=\alpha_D$.
Thus, for example, our model has the usual Ising critical
point at $\alpha_D=1.76$, and a reduced surface tension
$\gamma'=\gamma/kT$ given by Onsager's expression \cite{onsager44}
but with $\alpha_D$ replacing $\epsilon/kT$:
$\gamma'=\alpha_D/2-
\ln\left(\left[1+\exp(-\alpha_D/2\right]/
\left[1-\exp(-\alpha_D/2\right]\right)$.

The result 
of a simulation run at $\alpha_D=2.5$
is shown in
Fig.~\ref{fig:snaps}. We see that, as it must, it shows the usual
Ising lattice gas vapour/liquid coexistence.
We fix $r_M=8$, as varying it over a wide range of values (2 to 24)
is found to have little effect on the dynamics, and cannot change
the phase behaviour as varying $r_M$ does not alter the mapping
to the Ising lattice gas.

In terms of studying vapour/liquid coexistence in energy-consuming
systems,
the studies closest to this work are those of
the Edinburgh group
\cite{stenhammer13,wittkowski14,stenhammer14,stenhammer15,cates14,solon14}, who model
active Brownian particles with motilities that depend on the
local density of these particles, but no
attractions. Redner {\it et al.}~\cite{redner13} and
Mognetti {\it et al.}~\cite{mognetti13} also study systems with local slow down
in the dynamics that show condensation. In the Edinburgh group's work,
the particles condense into a
liquid phase coexisting with a dilute phase. Our models behave in a qualitatively
identical way, the fact that for the model here high densities decease
a hopping rate whereas it decreases a velocity in the model studied
in Edinburgh appears to make little difference.

However, our motivation is different. The Edinburgh group
focus on showing that a specific out-of-equilibrium model shows
equilibrium-like vapour-liquid phase separation, whereas we are interested
in constructing the simplest possible models model the liquid
droplets inside cells. There are also similarities between
our model, the models studied
by the Edinburgh group and others, and cooling granular media.
In cooling granular media locally high rates of inelastic collisions
in dense regions slow the particles in these dense
regions, which in turn causes clustering
that is partly analogous to condensation into a liquid-like
phase. See Paul and Das \cite{paul14} and references therein
for recent work in this area. However, there are differences,
these clusters are not at steady state and there
may well not be a well-defined surface tension there \cite{paul14}.


mRNA-rich droplets in the cytoplasm of cells are
not the only example of self-organised domains in cells.
There
is a huge amount of self-organised structure in the
chromatin in the nuclei of cells. Chromatin is the
DNA of our genes, together with associated proteins and RNA
\cite{misteli07,cope10,brangwynne11sm,sear15_cecam,mao11,brangwynne11nuc}.
Examples of this structure are domains called
nucleoli where ribosomes are made \cite{brangwynne11sm,mao11,brangwynne11nuc},
Cajal bodies where RNA splicing occurs \cite{brangwynne11sm,mao11},
and what is effectively microphase separation between
chromatin that is being transcribed and chromatin that is not
active \cite{mao11,misteli07,cope10,ganai14}.

So chromatin, which includes huge polymers of DNA,
exhibits behaviour reminiscent of the microphase separation
seen in block copolymers with immiscible blocks \cite{bates90}.
The nucleus is microphase separated in the sense that
lengths of chromatin that have a common feature, e.g., are making
ribosomal RNA in the case of nucleoli, have separated out from
the rest of the DNA.
However, the dynamics of the chromatin depends on energy consuming
processes \cite{sinha08,dundr07}, unlike in block
copolymers where the monomers move via thermal diffusion.
In conventional block copolymers the microphase separation is driven
by intermolecular attractions between like monomers
being stronger than between unlike monomers. The mechanism
may be different for chromatin.

Nuclear self-organisation is a very active field of
research \cite{misteli07,sear15_cecam,cope10,brangwynne11sm,mao11,brangwynne11nuc}
and we cannot
answer all the many unsolved questions here. However we can
see if density-dependent dynamics of the monomers of a block copolymer,
can result in microphase separation.
This would be a very simple model of separation between active and
inactive chromatin, or between a nucleolus and the surrounding
chromatin that is not involved in the making of ribosomes.

We introduce a simple out-of-equilibrium block
copolymer model.
The polymer consists of a linear chain
of $M$ monomers on a 2D square
lattice. The monomers are held together by bonds.
If the $x$ and $y$ coordinates of monomer $i$ are
$x_i$ and $y_i$ then the bonds along
the backbone are enforced by insisting that $|x_{i+1}-x_i|\le 1$
and $|y_{i+1}-y_{i}|\le 1$ for $i=1,M-1$.
The monomers are of two types: Type E which just interact via
excluded volume interactions, and type D which in addition have
dynamics that depend on the local density.

\begin{figure}[tb!]
\includegraphics[width=5.5cm,angle=0]{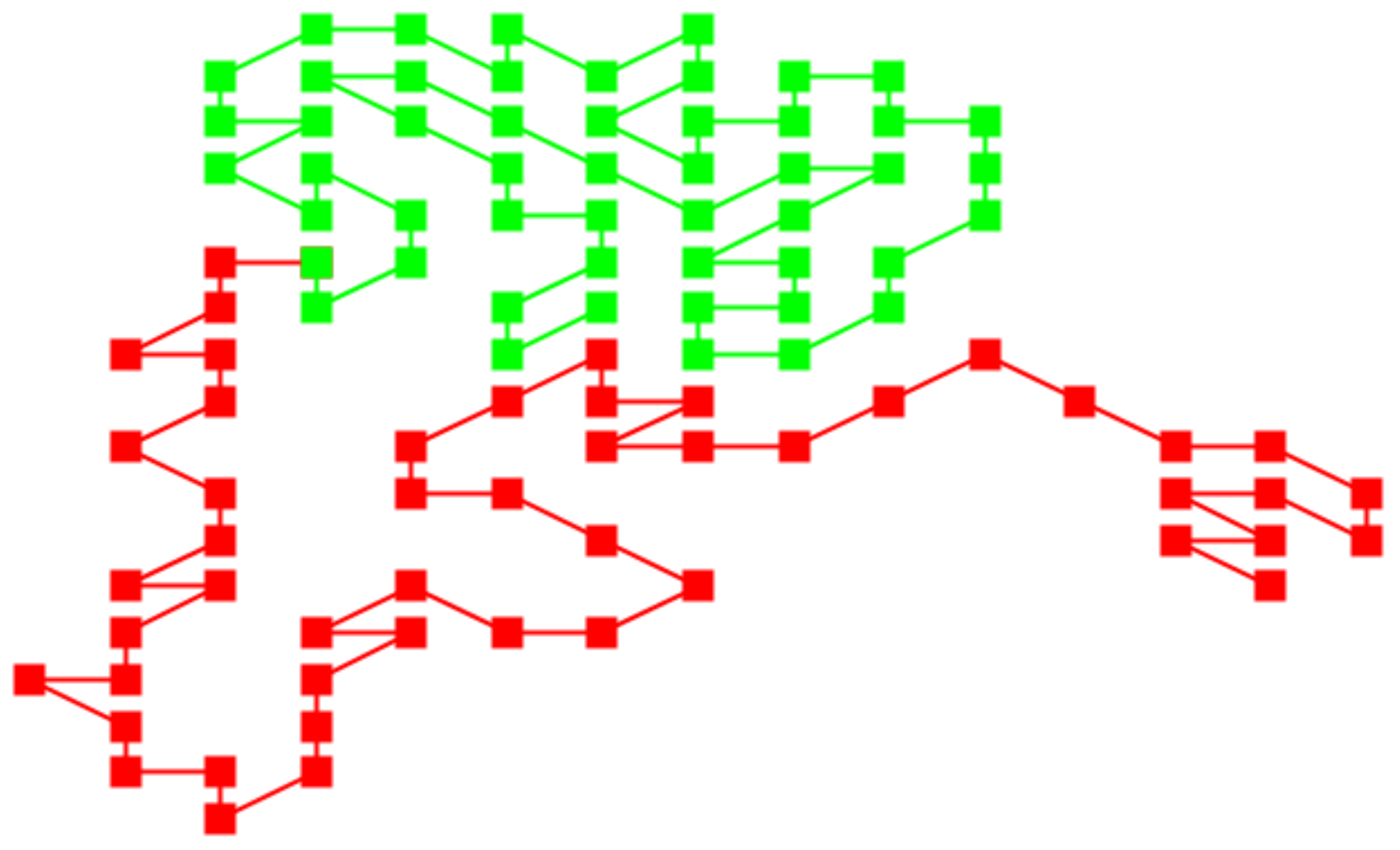}
\caption{Snapshot of a configuration of a 2D lattice block copolymer
of length $M=100$, with 50 of each monomer type.
The monomers in green (type D) have dynamics
that slow in the presence of other monomers, whereas those in
red (type E) interact only via excluded volume.
The parameter $\alpha_{DP}=0.6$.}
\label{fig:polymer}
\end{figure}

A simulation cycle consists of $M$ attempted moves, in each of
which one of the $M$ monomers is selected and an attempt
is made to move it to one of the eight neighbouring lattice sites.
These eight are the sites above, above right, right, below right, etc,
of the central site.
If a type E monomer has been selected, that move is always successful
unless it would move the monomer to an already occupied site, in which
case it is always rejected.
If a type D monomer is selected then the move is also rejected if
it would move the monomer to an already occupied site.
But if the site is vacant, the move is only made with a probability
$p=\exp[-\alpha_{DP}n_{MN}]$.
Here $n_{MN}$ is the total number of monomers
(excluding the monomers it is bonded to)
in the
eight lattice sites that surround the site the molecule would be moved from.
The parameter $\alpha_{DP}$ controls the density dependence of the dynamics.
The polymer is simulated on a large lattice; periodic boundary conditions
are not used.
Note that this model for density dependent dynamics also
satisfies detailed balance, for the same reason as for our monomeric
model. The model has hard constraints, both restricting the molecules
to no more than one per site and restricting the bond length between
successive monomers. For two states $i$ and $j$ that are allowed
as they satisfy these requirements the transition
probabilities satisfy
$p_{ij}/p_{ji}=\exp\left[-\alpha_{DP}[ n_{MN}(i)- n_{MN}(j)]\right]$,
and so satisfy detailed balance.

For a block copolymer of the two types of monomer, the two halves
of the polymer separate, with the type D monomers
forming a condensed globule, while the type E monomers
form an extended chain. A snapshot of this is shown in Fig.~\ref{fig:polymer}.

Note that our model for microphase separation
is different to Ganai {\it et al.}~\cite{ganai14}'s
model for separation of chromatin
that is being transcribed from chromatin that is not being transcribed. In their
work domain formation is driven by a large increase in effective
temperature in chromatin that is being transcribed, without
a direct mechanism for density-dependence of the dynamics.
Within our model, transcriptionally active regions of chromatin
separate if the active dynamics of these regions are such that
their motion slows when in contact with other active regions.
Similarly, applied to a nucleolus, our model
predicts that the nucleolus separates from the surrounding chromatin
due to ribosome-producing chromatin slowing its mesoscale
motion in the presence
of other ribosome-producing chromatin --- perhaps because it is sharing
factors or due to direct attractive interactions.


We have studied very simple models and found that molecules whose hopping rate decreases strongly when the
local density is high, condense into liquid
droplets. The simplicity of our model, and the fact that
condensation is also seen in
very different models in which the velocity decreases
when the local density is high\cite{stenhammer13,wittkowski14,cates14,mognetti13,redner13},
suggests that this condensation is quite generic.
The model is an equilibrium one in the sense that
it obeys detailed balance, but is for out-of-equilibrium
systems that consume energy. Thus to the extent that
it models the microscopic dynamics of mRNA molecules in cells correctly,
their behaviour will be qualitatively that of an equilibrium liquid.
Although the model maps to the Ising lattice gas, an equilibrium
model, the energy driving the dynamics does not have to be $kT$.
For mRNA molecules in cells, a better candidate, at least for
long distance motion \cite{fusco03}, is the
much larger forces that molecular
motors can provide. This has consequences. For example
the effective surface tension that limits the fluctuations
of the liquid surface
is then not $\sim kT/a^2$,
it is presumably $\sim w_M/a^2$, with $w_M$ the work done by
one or a few motors on one of the molecules of the liquid. Here
$a$ is a molecular diameter.

\section*{Acknowledgements}

It is a pleasure to acknowledge Steve Whitelam for pointing
out the mapping to the Ising model, and
helpful discussions with Ignacio Pagonabarraga and Daan Frenkel.


%

\end{document}